\documentclass[aps,prb,letterpaper,notitlepage,longbibliography,reprint]{revtex4-1}
\newcommand{\Smat}{\mathcal{S}}
\newcommand{\Tmat}{\mathcal{T}}

\usepackage{xcolor,graphicx,amsmath,amsfonts,amssymb,amsthm,bm,bbm}
\usepackage{ifthen}

\usepackage[pdfpagelabels,pdftex,bookmarks,breaklinks]{hyperref}
\definecolor{darkblue}{RGB}{0,0,127} 
\definecolor{darkgreen}{RGB}{0,180,0}
\hypersetup{
colorlinks,
linkcolor=darkblue,
citecolor=darkgreen,
filecolor=red,
urlcolor=blue,
pdftitle={Tensor Networks with a Twist},
pdfauthor={Jacob C. Bridgeman, Stephen D. Bartlett, Andrew C. Doherty}
}

\definecolor{tctwistcolor}{RGB}{255,0,0}
\definecolor{tcmcolor}{RGB}{0,0,180}
\definecolor{tcecolor}{RGB}{0,180,0}
\definecolor{tcTTppcolor}{RGB}{184,134,11}
\definecolor{tcTTpmcolor}{RGB}{252,15,192}

\newcommand{\ZZ}{\mathbb{Z}}

\newcommand{\Fref}[1]{Fig.~\ref{#1}}

\let\oldonlinecite\onlinecite
\renewcommand{\onlinecite}[1]{Ref.~[\oldonlinecite{#1}]}

\DeclareMathOperator{\Di}{Dih}

\newcommand{\ket}[1]{|{#1}\rangle}

\newcommand{\ketbra}[2]{|{#1}\rangle\!\langle{#2}|}

\newcommand{\G}{G}
\newcommand{\Dih}[1]{\Di_{#1}}

\newcommand*{\one}{\mathbbm{1}} 
\newcommand{\D}[1]{\mathcal{D}{} }

\newcommand{\restrict}[1]{\raise-.2ex\hbox{\ensuremath|}_{#1}}

\definecolor{tensorblue}{rgb}{0.8,0.8,1}
\definecolor{tensorred}{rgb}{1,0.5,0.5}
\definecolor{tensorgreen}{rgb}{0.6,1,0.6}
\definecolor{tensorpurp}{rgb}{1,0.5,1}

\makeatletter
\newcommand{\vast}{\bBigg@{4}}
\newcommand{\Vast}{\bBigg@{9}}
\makeatother

\def\Put(#1,#2)#3{\leavevmode\makebox(0,0){\put(#1,#2){#3}}}

\makeatletter
\def\pgf@plot@curveto@handler@finish{%
\ifpgf@plot@started%
\pgfpathcurvebetweentimecontinue{0}{0.995}{\pgf@plot@curveto@first}{\pgf@plot@curveto@first@support}{\pgf@plot@curveto@second}{\pgf@plot@curveto@second}%
\fi%
}
\makeatother

\newlength\figureheight
\newlength\figurewidth

\newcommand{\includeTikz}[2]{\includegraphics{figures/#1}
}

\definecolor{nicegreena}{RGB}{1,115,16}
\definecolor{nicegreenb}{RGB}{1,240,16}

\definecolor{nicegreen}{RGB}{60,183,82}

\colorlet{ccred}{red!20}
\colorlet{ccgreen}{green!50}
\colorlet{ccblue}{blue!20}

\definecolor{tensorblue}{rgb}{0.8,0.8,1}
\definecolor{tensorred}{rgb}{1,0.5,0.5}
\definecolor{tensorpurp}{rgb}{1,0.5,1}

\usepackage{soul}
\newcommand{\add}[1]{#1}
\newcommand{\del}[1]{}

\allowdisplaybreaks
\begin{document}

\title{Tensor Networks with a Twist:\\Anyon-permuting domain walls and defects in PEPS}

\author{Jacob C.\ Bridgeman}
\author{Stephen D.\ Bartlett}
\author{Andrew C.\ Doherty}
\affiliation{Centre for Engineered Quantum Systems, School of Physics, The University of Sydney, Sydney, Australia}

\date{\today}

\begin{abstract}
We study the realization of anyon-permuting symmetries of topological phases on the lattice using tensor networks. Working on the virtual level of a projected entangled pair state, we find matrix product operators (MPOs) that realize all unitary topological symmetries for the toric and color codes. These operators act as domain walls that enact the symmetry transformation on anyons as they cross. By considering open boundary conditions for these domain wall MPOs, we show how to introduce symmetry twists and defect lines into the state.
\end{abstract}

\maketitle

The low energy states of strongly interacting spin models can exhibit complex and exotic physics. A particularly interesting class of models are those that are topologically ordered\cite{WEN1990a,Nayak2008,Wen2013}. The ground spaces of these models are promising candidates for robust storage of quantum information\cite{Dennis2002,Kitaev2003,Brown2014,Terhal2015,Brell2014}.

If quantum information is encoded in the degenerate ground space of topologically ordered systems, the action of anyon-permuting symmetries (APS) can be used to apply logical transformations\cite{Dennis2002}. These symmetries map among quasi-particle excitations without changing the topological phase. Large classes of symmetry actions give the potential for fault-tolerant logic manipulation, a prerequisite for effective quantum computation. Additionally, the introduction of symmetry defects can increase the functionality of the code for quantum computation\cite{Brown2016}. It is therefore important to understand the interplay of symmetry and topological order in such spin models\cite{Etingof2010,Kitaev2006,Barkeshli2013,Barkeshli2014,Tarantino2016,Heinrich2016,Cheng2016}.

Recently, a connection has been made between fault-tolerant logical gates, locality-preserving symmetries and anyon-permuting domain walls. In particular, an  equivalence was established between such logical gates and domain walls for topological stabilizer codes\cite{Beverland2014,Yoshida2015a,Yoshida2017,Webster}.

In this paper, we take this connection as our starting point and investigate realizations of such domain walls in two-dimensional topologically ordered models using projected entangled pair states (PEPS) and matrix product operators (MPOs)\cite{Verstraete2008,Pirvu2010,Orus2014,Bridgeman2017}. These tools allow the efficient representation of ground states of topologically ordered models\cite{Schuch2010a,Schuch2013,Schwarz2013,Sahinoglu2014a,Bultinck2017}, and provide a useful framework for the construction of domain walls.

By working with two important examples, the toric and color codes, we show how to construct the domain walls corresponding to all APS and investigate their properties. In particular, we do this without modifying the underlying PEPS description of the state.
We are further able to construct states containing APS defects\cite{Bombin2010,Kitaev2012,Brown2013,Barkeshli2013,Barkeshli2014,Teo2015a,Teo2015,Tarantino2016}.
These defects enrich the properties of the underlying topological model. In particular, they may allow more exotic fusion and braiding than the original anyons\cite{Bombin2010,Barkeshli2013,Barkeshli2014}, which can lead to increased computational power within a model of topological quantum computing. The defects can also be used to introduce additional encoded qubits, increasing the storage capacity of a code\cite{Brown2016}.

Although our discussion centers around two exactly solvable spin systems, the framework we are advocating should be far more general. PEPS make it straightforward to move away from simple fixed point models (models with zero correlation length). On the physical lattice, one expects the string operators associated to the anyons to `spread out' into wider ribbons\cite{Hastings2005,Bridgeman2016}, which makes the local action of the APS operators much more complicated, whilst on the virtual level of the PEPS these anyon string operators remain fully localized\cite{Bultinck2017}.

This paper is organized as follows: In Section~\ref{S:review} we review some ideas important to this work, and introduce some notation for the remainder of the paper. In Section~\ref{S:toriccode} we introduce the toric code, including the anyon-permuting symmetries.
We then introduce a PEPS for the ground states of this model and discuss the realization of an anyon-permuting domain wall on this PEPS. Finally, we show how to introduce APS defects carrying a definite generalized charge, and discuss fusion and parent Hamiltonians of such defects. In Section~\ref{S:colorcode} we discuss the color code, a topological model with far richer symmetries than the toric code. We construct all domain walls of this topological phase, and the corresponding defects. In Section~\ref{S:conclusions} we summarize the results and discuss possible extensions. For completeness, we include stabilizers for topological states with symmetry twists in Appendix~\ref{S:twisthamiltonians}, and construct the domain wall MPOs for the $\ZZ_N$ generalizations of the toric code in Appendix~\ref{S:quantumdoublemodels}.

\section{Review: Topological order and PEPS}\label{S:review}

In this section, we review some key concepts, notation and conventions required for the remainder of the paper.

We begin with a discussion of topologically ordered phases, the kind of symmetries they support and the connections to fault-tolerant quantum computation. This motivates the discussion of locality preserving APS actions, domain walls, and defects. These topics form the primary objects of study in this paper.

We introduce PEPS, the main tool used in this work, and a streamlined notation we use throughout the paper. Following this, we discuss how local symmetries can be realized in PEPS for systems without topological order. This motivates our realization of APS domain walls using matrix product operators. We then describe topologically ordered PEPS, which form the basis of the remainder of the paper.

\subsection{Topological order and anyon-permuting symmetries}\label{S:symmetries}

For our purposes, an intrinsic topological phase is defined by a set of anyon labels $\{a_i\}$ and their braiding and fusion rules. 
These quasi-particles are a generalization of bosons and fermions, and can exhibit more complex braid relations. These relations are captured by the $\Tmat$ and $\Smat$ matrices of the theory
\begin{align}
\Tmat_{a} = \frac{1}{d_a}\!\!
\begin{array}{c}
\includeTikz{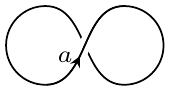}{}
\end{array}\!,\,\,\,
\Smat_{a,b}=\frac{1}{D}\!\!
\begin{array}{c}
\includeTikz{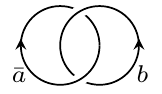}{}
\end{array}
\!,\label{eqn:STmatrix}
\end{align}
where $d_a$ is the quantum dimension of anyon $a$ and $D=\sqrt{\sum_a d_a^2}$ is the total quantum dimension. These matrices define the self- and mutual- braid relations of the particles respectively\cite{Kitaev2006}, and implicitly define the fusion rules.


We consider symmetries of the topological phase corresponding to a permutation of the anyon labels that preserves these braiding relations\cite{Tarantino2016,Kitaev2012,Teo2015a,Teo2015,Brown2013,Barkeshli2013,Heinrich2016,Cheng2016,Bombin2010,Yoshida2015a}. We call this an anyon-permuting symmetry (APS). As a consequence, the fusion rules are also preserved. In particular, this means the vacuum must be invariant under any APS.

An anyon model can arise as the low energy spectrum of a gapped many-body spin model.
On the microscopic spin model, the APS may be realized via some complicated operator. The essential features, however, are captured by the action on the emergent quasi-particles
\footnote{A given spin model may break the APS, leading to a richer theory of symmetry enriched topological phases\cite{Barkeshli2013,Barkeshli2014,Cheng2016,Heinrich2016,Tarantino2016,Teo2015}. We will focus on models that do not break the symmetry. }.

In general, the action of an APS has no locality constraints, however it is natural to assume they will respect the underlying locality of the model.
This means that the action of the symmetry should map local operators to local operators. The most simple example is a transversal (on-site) action, but more generally the APS may be realized as finite depth quantum circuits and spatial transformations such as translations.

\subsubsection{Domain walls}

\begin{figure}
\centering
\includeTikz{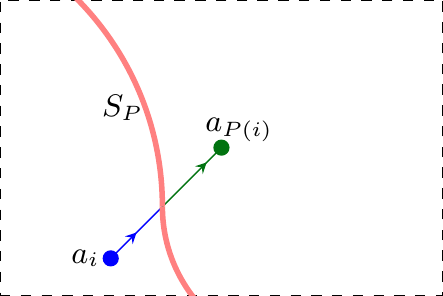}{}
\caption{Domain walls $S_P$ induce a permutation $P$ on the set of anyons $\{a_i\}$ when they cross. To ensure the topological phase is the same on both sides of the wall, only permutations that preserve the matrices $\Smat$ and $\Tmat$ of Eqn.~\ref{eqn:STmatrix} are allowed.}\label{fig:permutingDW}
\end{figure}

When a locality preserving APS acts on some region $\mathcal{R}$ of the lattice, it must act trivially far away from the boundary of $\mathcal{R}$. This is because, from the point of view of any operator within $\mathcal{R}$ with support far from the boundary, the symmetry has been applied to the entire system. If the state to which the symmetry was applied was a ground state, far from the boundary the state still looks locally like a ground state. Conversely, if there was an anyon of type $a_i$ within the region $\mathcal{R}$, it is transformed to an anyon of type $a_{P(i)}$, where $P$ is some permutation.
The symmetry operator can then be identified with a transparent domain wall in the vicinity of the boundary\cite{Yoshida2015a} of $\mathcal{R}$. Anyons are transformed when they cross such a domain wall as shown in Fig.~\ref{fig:permutingDW}.

\subsubsection{Topological order and fault-tolerant quantum computation}

Topologically ordered models are of great interest in quantum information theory. Logical information can be encoded in the degenerate ground space of the model\cite{Dennis2002,Kitaev2003,Brown2014,Terhal2015,Brell2014}, and is protected from local noise processes by the topological properties of the model. A logical gate is any transformation that maps among the allowed logical states. In the case of a topological code this is any transformation that preserves the ground space. Since any APS preserves the ground space of a topologically ordered spin model, the action of such symmetries can be used to enact logical operations.
Locality preserving APS are particularly interesting from a quantum information perspective. This kind of locality preservation means that the action does not spread errors in the code to the point where encoded information is corrupted, and is referred to as a fault-tolerant logical gate\cite{Brown2014,Terhal2015}. A key question in quantum information is how to identify sets of fault-tolerant logic gates for a given quantum error correcting code\cite{Eastin2008,Bombin2011,Bravyi2013a,Kubica2015}. The identification of anyon-permuting domain walls attempts to address this question for topological codes\cite{Yoshida2015a,Yoshida2017}.

\subsection{Projected entangled pair states}\label{S:PEPSreview}

In this section we review some of the key properties of PEPS representations of symmetric states, and some aspects of topologically ordered PEPS. For simplicity, we will assume translation invariance, although this is not crucial.

A PEPS representation of a state is described using a set of tensors $A$, which we represent as
\begin{align}
A_{\alpha,\beta,\gamma,\delta}^{i}&=
\begin{array}{c}
\includeTikz{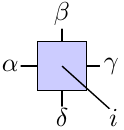}{}
\end{array},
\end{align}
where the greek indices are referred to as virtual, and the roman index is `physical'. A PEPS corresponds to a network of these tensors
\begin{align}
\ket{\psi[A]}&=
\begin{array}{c}
\includeTikz{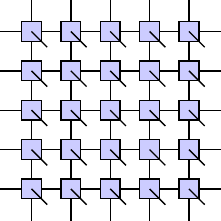}{}
\end{array},
\end{align}
where a line joined to a pair of tensors indicates contraction of indices, and some choice of boundary conditions should be chosen. For a review of tensor network notation and PEPS, we refer the reader to \onlinecite{Bridgeman2017}. A tensor may have more than one physical index attached to it, and we will usually neglect drawing these to simplify the diagrams. Frequently, we will also suppress drawing the tensors themselves. They will be implied at the intersection of indices. The notation for the above state will therefore be
\begin{align}
\ket{\psi[A]}&=
\begin{array}{c}
\includeTikz{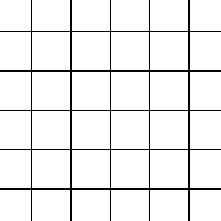}{}
\end{array}.
\end{align}

We now review the inclusion of local, physical symmetries in PEPS without intrinsic topological order, and the inclusion of topological order via a virtual symmetry. In Section~\ref{S:toriccode}, we show how these two properties can be combined.

\subsubsection{Local symmetries in PEPS}\label{S:symmetrypeps}

Consider, for the moment, the class of PEPS describing ground states of systems with no topological order, and no spontaneous breaking of the symmetry.
Within this class of PEPS, the action of a transversal symmetry on a region $\mathcal{R}$ can be realized by an MPO domain wall acting on the virtual bonds around the edge\cite{Williamson2016} of $\mathcal{R}$
\begin{align}
\begin{array}{c}
\includeTikz{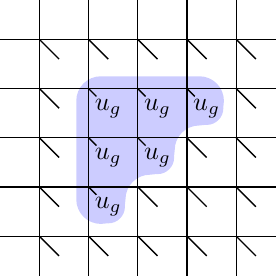}{}
\end{array}
&=
\begin{array}{c}
\includeTikz{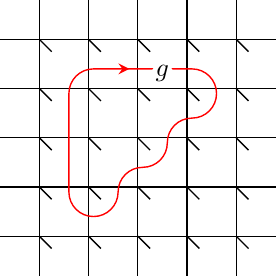}{}
\end{array},
\end{align}
where the MPO tensors occur at the intersection of a red and black line. The four-index MPO tensors have two (black) indices acting on the virtual bonds of the PEPS, and two (red) `virtual' indices. These virtual indices are contracted to give the operator. The MPOs are labelled by a group element $g$, and the collection of MPOs forms a representation of the symmetry group, so
\begin{align}
\begin{array}{c}
\includeTikz{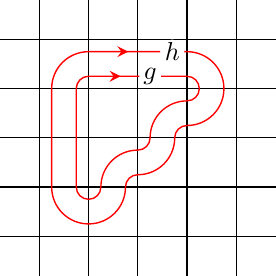}{}
\end{array}
&=
\begin{array}{c}
\includeTikz{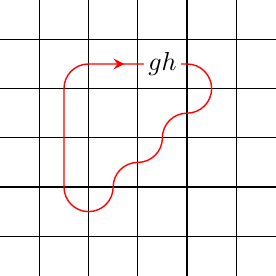}{}
\end{array}.\label{eqn:mporep}
\end{align}

The MPO can be pulled through the PEPS, leaving behind the physical symmetry action
\begin{align}
\begin{array}{c}
\includeTikz{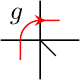}{}
\end{array}
&=
\begin{array}{c}
\includeTikz{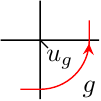}{}
\end{array},
\end{align}
thereby allowing the symmetry transformed domain to be enlarged. In this way, we can propagate the domain wall across the lattice, at the expense of a physical action. This `pulling through' condition ensures the domain wall/symmetry correspondence holds on all regions.

This framework of virtual MPO representations allows for all symmetry protected topological phases with on-site symmetry action to be realized in PEPS\cite{Williamson2016}. As discussed in Section~\ref{S:topopeps}, a similar framework allows for the construction of PEPS with intrinsic topological order but no symmetry. The aim of this paper is to combine these two properties in familiar PEPS states, without altering the underlying topologically ordered state.

\subsubsection{Topologically ordered PEPS}\label{S:topopeps}

In this section, we briefly review a class of PEPS supporting intrinsic topological order. We restrict our discussion to $\G$-injective PEPS, which describe the ground states of quantum double models\cite{Schuch2010a}.

Unlike the PEPS discussed in Section~\ref{S:symmetrypeps}, a $\G$-injective PEPS does not necessarily have any physical symmetry, but does support a virtual symmetry
\begin{align}
\begin{array}{c}
\includeTikz{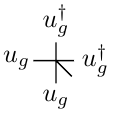}{}
\end{array}
&=
\begin{array}{c}
\includeTikz{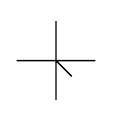}{}
\end{array}.\label{eqn:ginj}
\end{align}
\del{when sufficiently many PEPS tensors have been blocked together.} \add{For PEPS with nontrivial correlation length (i.e. away from renormalization group fixed points) Eqn.~\ref{eqn:ginj} may only hold on renormalized tensors obtained by contracting patches of the original PEPS tensors together.} This is not a gauge transformation as it holds at the single tensor level rather than $u$ and $u^\dagger$ being applied to adjacent tensors. This condition is encoding a version of Gauss' law for the \add{topological charges (anyons)\cite{LGT}}.

The PEPS is said to be injective if the tensor realizes an injective map from the virtual to physical indices. A PEPS with a virtual symmetry cannot be injective (if the representation $u_g$ is not the trivial representation), but can be $\G$-injective, meaning the map is injective on the $\G$-invariant subspace.

Anyons can be represented on the virtual level of the PEPS\cite{Schuch2010a,Haegeman2014,Duivenvoorden2017} . Magnetic particles, labelled by a conjugacy class containing $g$, can be inserted into the PEPS by inserting open strings of $u_g$ on the virtual level
. The bulk of these strings can be deformed using Eqn.~\ref{eqn:ginj}, but the end points are pinned in place and can therefore be measured. Electric ($e$) charges, labelled by an irreducible representation $\chi$, can also be included. For simplicity, assume that $\G$ is abelian. An $e$ particle can then be inserted using an operator $X_\chi$ on the virtual level such that $X_\chi u_g=\chi(g) u_g X_\chi$, which ensures the correct braiding relations. We remark that the insertion of an electric particle is not associated to a string. One can therefore define a state with a single $e$, but there is no physical operator that can construct such a state. We will describe this for the special case $\G=\ZZ_2$ in Section~\ref{S:toriccode}.

\section{Anyon-permuting symmetries of the toric code}\label{S:toriccode}

In this section, we find an MPO that realizes the $\ZZ_2$ APS of the simplest topological phase, the toric code\cite{Kitaev2003}. This model describes a phase with a topological order known as the $\ZZ_2$ quantum double. We will first introduce the model and a PEPS realizing this topological order.

The topological phase is defined by four anyons, conventionally labelled $\{1,e,m,em\}$. The fusion rules are $a\times a=1$ for all anyons and $e\times m=em$. The only nontrivial element of the $\Tmat$ matrix is $\Tmat_{em}=-1$, whilst all nontrivial $\Smat$ matrix elements are obtained from  $\Smat_{e,m}=-1/2$. The only APS is a $\ZZ_2$ symmetry defined by the action $\D{}(e)=m$ and $\D{}(e)=m$. Note that this APS is distinct from the $\ZZ_2$ symmetry defining the phase, which preserves the charge mod 2.

\subsection{Topological PEPS}

\begin{figure}
\centering
\includeTikz{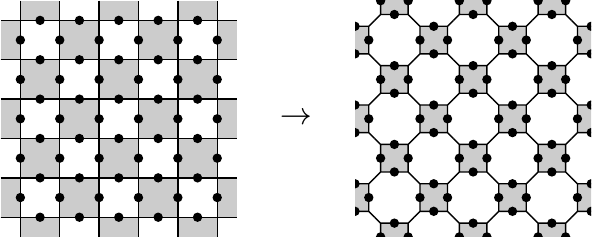}{}
\caption{The toric code is defined on a square lattice, with spins on edges. It is convenient to write a PEPS tensor for each shaded plaquette. By inserting a bond at the vertex connecting adjacent gray plaquettes, a 4.8.8 lattice is obtained.
}\label{fig:tclattice}
\end{figure}

To construct the minimal square lattice PEPS realizing the $\ZZ_2$ quantum double topological order, we consider Eqn.~\ref{eqn:ginj} with $u_g=Z$, the qubit Pauli $Z$ operator. To construct a $\ZZ_2$-injective PEPS with this symmetry, we must ensure the dimension of the PEPS tensor in Eqn.~\ref{eqn:ginj} is at least $2^4/2$, since there are four virtual bonds each with dimension 2 and half of the virtual space is symmetric (the even-parity subspace). It is therefore convenient to construct a PEPS tensor for each shaded plaquette in Fig.~\ref{fig:tclattice}, so that there are four physical qubits per tensor.

Due to the topological order, the tensor must have a local, virtual $\ZZ_2$ symmetry
\begin{align}
\begin{array}{c}
\includeTikz{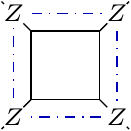}{}
\end{array}
&=
\begin{array}{c}
\includeTikz{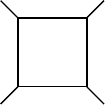}{}
\end{array},\label{eqn:tctoposymmetry}
\end{align}
and must be an injective map (on the $\ZZ_2$-invariant subspace) from virtual to physical indices\cite{Schuch2010a}.
It is straightforward to check that a PEPS with nonzero elements
\begin{align}
\begin{array}{c}
\includeTikz{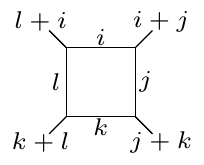}{}
\end{array}
=1,\label{eqn:tcpepstensor}
\end{align}
has this symmetry. Here, we place a physical spin on each horizontal/vertical edge, diagonal edges correspond to the virtual indices, and all additions are taken modulo 2.

Anyons can be represented directly on the virtual bonds of the PEPS. As discussed in Section~\ref{S:PEPSreview}, a state with a pair of $m$ particles is created by
\begin{align}
\begin{array}{c}
\includeTikz{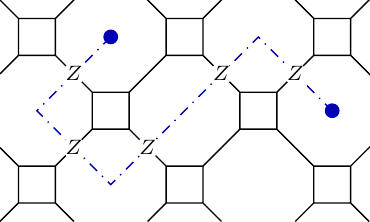}{}
\end{array},
\end{align}
where the path of the string is arbitrary since the virtual symmetry Eqn.~\ref{eqn:tctoposymmetry} can be used to move it. A state with two $e$ anyons is created by
\begin{align}
\begin{array}{c}
\includeTikz{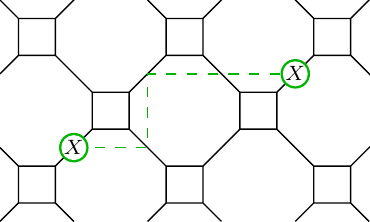}{}
\end{array},\label{eqn:tcee}
\end{align}
where the $X$ operators are only placed at the end points of the string, corresponding to the location of the excitations. Unlike the $m$ type anyons, there is no string associated to the $e$ particles on the virtual level of the PEPS.
In this sense, the $e$ anyons are `localized' since the presence of an $X$ operator signals the location of a particle. On the other hand, a $Z$ does not signal the location of an $m$ since strings of $Z$ operators can be fluctuated through the PEPS using Eqn.~\ref{eqn:tctoposymmetry}. It is therefore possible to define a single $e$ particle, by inserting a single $X$ operator, although there is no operation on the physical bonds that creates such a state. Conversely, no state with a single $m$ excitation can be defined.

\subsection{Anyon-permuting symmetry}
The only APS of this model is the transformation $e\leftrightarrow m$. On the virtual level, this can be implemented by an operator that transforms pairs of $X$ operators to strings of $Z$s. We recognize this transformation as the Ising duality map $\D{}^{(o)}$
\begin{align}
\D{}^{(o)}{}^\dagger X_j \D{}^{(o)}&= \prod_{k\leq j}Z_k\\
\D{}^{(o)}{}^\dagger Z_j \D{}^{(o)}&= X_j X_{j+1}\label{eqn:Isingdualityb}
\end{align}
performs \del{has }the desired action. On a line, this can be implemented by the circuit
\begin{align}
\D{}^{(o)}&=
\begin{array}{c}
\includeTikz{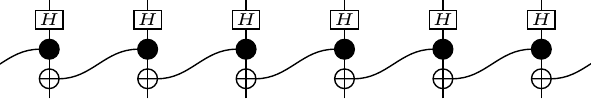}{}
\end{array},
\end{align}
where $H$ is the Hadamard operator and $\includeTikz{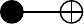}{}$ is the controlled-$X$ operator.

Since domain walls act around closed paths, corresponding to the boundary of a domain of symmetry action, it is important to define a periodic version of this circuit. This can be done by noting that the circuit $\D{}$ is realized by the MPO
\begin{align}
\D{}^{(o)}&=
\begin{array}{c}
\includeTikz{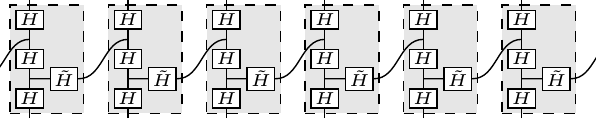}{}
\end{array},\label{eqn:do}
\end{align}
where $\tilde{H}=\sqrt{2}H$, and
\begin{align}
\begin{array}{c}
\includeTikz{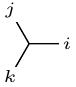}{}
\end{array}
&=\begin{cases}
1\qquad\text{if $i=j=k$}\\
0\qquad\text{otherwise}
\end{cases}.
\end{align}
The virtual indices of this MPO can be connected to produce a periodic operator that will be referred to as $\D{}$. The translationally invariant domain wall MPO is defined by
\begin{align}
\begin{array}{c}
\includeTikz{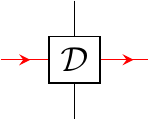}{}
\end{array}
&=
\begin{array}{c}
\includeTikz{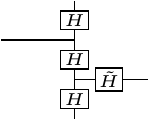}{}
\end{array},\label{eqn:TCWall}
\end{align}
where the arrow indicates that in Eqn.~\ref{eqn:Isingdualityb} we made a choice $Z_j\to X_jX_{j+1}$ rather than $Z_j\to X_{j-1}X_j$.

Since this MPO tensor is injective (as a map from virtual to physical indices), the tensor is unique up to gauge transformations\cite{Haegeman2017}. One such gauge transformation is the choice to block $\tilde{H}$ on the right, rather than the left, of the tensor in Eqn.~\ref{eqn:do}.

On periodic boundaries, this operator ceases to be unitary, but remains an isometry up to a rescaling of $1/\sqrt{2}$. This can be seen by noting that
\begin{align}
\begin{array}{c}
\includeTikz{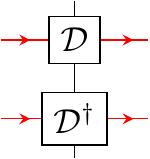}{}
\end{array}
&=
\begin{array}{c}
\includeTikz{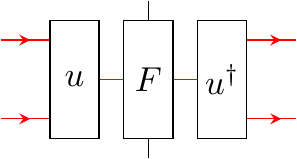}{}
\end{array},
\end{align}
where
\begin{align}
\begin{array}{c}
\includeTikz{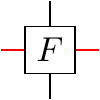}{}
\end{array}
&=
\begin{pmatrix}
\one&0&0&0\\
0&Z&0&0\\
X&0&0&0\\
0&XZ&0&0
\end{pmatrix}
,
\end{align}
and $u$ is a unitary gauge transformation.
The notation here identifies left (right) virtual indices to row (column) indices of the matrix and up (down) physical indices (black) correspond to operator indices of matrix entries.

The matrix $u$ is unitary, so on periodic MPOs, these cancel with $u^\dagger$ from the neighboring tensor. The off-diagonal elements of $F$ do not contribute on periodic MPOs since the trace (with respect to the virtual indices) is taken. The MPO on $N$ sites is therefore $\one^{\otimes N}+Z^{\otimes N}$, corresponding to (twice) the projector onto the even parity subspace.
On this subspace, which corresponds to the support of the PEPS, $\D{}$ is unitary. We remark that the wall defined by $\D{}^\dagger$ permutes the set of anyons in the same way as $\D{}$, so corresponds to the same topological symmetry action.

By using the representation of anyons on the PEPS, along with the $\ZZ_2$ APS, we have constructed an explicit MPO realizing the APS. For the remainder of this section we describe the properties of this domain wall MPO, including how to terminate open walls to create APS twists with definite generalized topological charge.

\subsubsection{Algebra of domain walls}\label{ss:growing}

\begin{figure}
\centering
\includeTikz{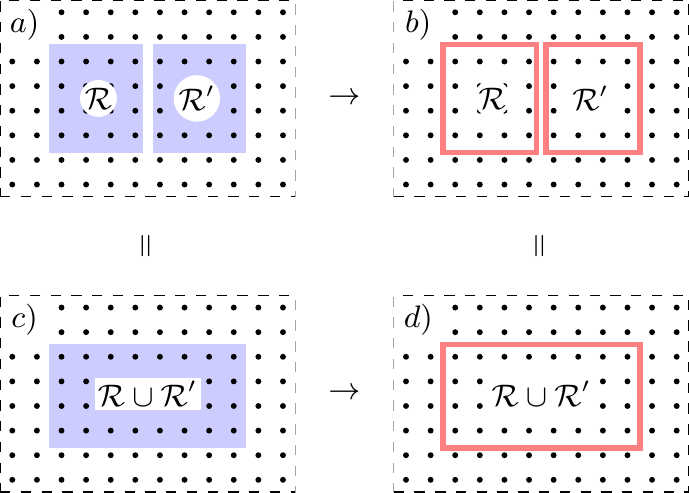}{}
\caption{Applying a transversal (on-site) symmetry to regions $\mathcal{R}$ and $\mathcal{R}^\prime$ (a) corresponds to domain walls at the boundaries $\partial \mathcal{R}$ and $\partial \mathcal{R}^\prime$ (b). If the symmetry is transversal, is the same as applying the symmetry to $\mathcal{R}\cup\mathcal{R}^\prime$ (c), so the domain walls should merge (d). For a non-transversal symmetry, a nontrivial operator will remain along the merge.}\label{fig:transversal}
\end{figure}

It is important to understand the algebra of the domain walls so that the action of multiple walls can be computed. As an example, we will study the effect of applying the APS operator to disjoint regions of the lattice.
The MPOs form a representation \add{of} the APS group when multiplied along their whole length, as in Eqn.~\ref{eqn:mporep}, but not at the local tensor level. Therefore, the multiplication of domain walls on regions such as that in Fig.~\ref{fig:transversal}b) cannot be deduced directly from the group multiplication.


Consider applying an APS to adjacent regions $\mathcal{R}$ and $\mathcal{R}^\prime$ as depicted in Fig.~\ref{fig:transversal}. This corresponds to applying domain walls around the two boundaries $\partial\mathcal{R}$ and $\partial\mathcal{R}^\prime$. For a transversal (on-site) action, the action on the full region $\mathcal{R}\cup\mathcal{R}^\prime$, so the domain walls should merge, leaving a single wall around $\partial(\mathcal{R}\cup\mathcal{R}^\prime)$. For a non-transversal action, corresponding to a finite depth circuit, acting on $\mathcal{R}$ and $\mathcal{R}^\prime$ separately is not equivalent to acting on $\mathcal{R}\cup\mathcal{R}^\prime$ (i.e. Fig.~\ref{fig:transversal}a and  Fig.~\ref{fig:transversal}c are not equal). There are missing gates along the shared boundary.

\begin{figure}
\centering
\includeTikz{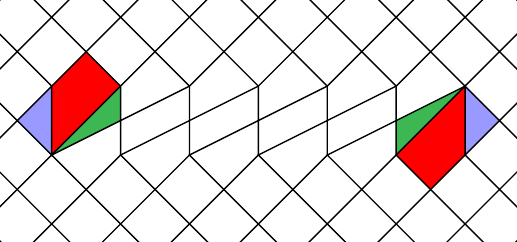}{}
\caption{When merging two domain walls, the lattice geometry is altered along the wall. This is because the symmetry does not act transversally. Along the interface, a lattice translation is implemented. The colors correspond to those in the tensor network of Eqn.~\ref{eqn:translationdefect}.}
\label{fig:tctranslationdefect}
\end{figure}

We study this merging effect using the domain wall constructed in Eqn.~\ref{eqn:TCWall}. To determine the difference between Fig.~\ref{fig:transversal}b and Fig.~\ref{fig:transversal}d, we proceed by applying the inverse of the larger wall (d), followed by the action on the smaller regions (b). If the APS is realized transversally these actions will cancel out, but if the symmetry is merely locality preserving there will be an action along the `join'. Denoting $\D{}^\dagger$ by a blue line and $\D{}$ by a red line, the action on the PEPS is
\begin{align}
\begin{array}{c}
\includeTikz{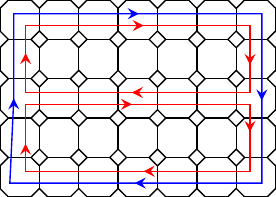}{}
\end{array}
&=\sum_{i,j=0}^1
\begin{array}{c}
\includeTikz{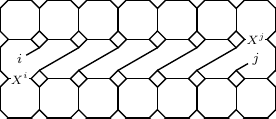}{}
\end{array}\\
&=
\begin{array}{c}
\includeTikz{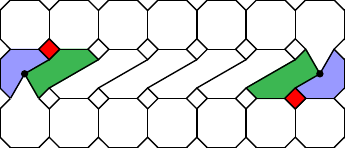}{}
\end{array},\label{eqn:translationdefect}
\end{align}
where
\begin{align}
\begin{array}{c}
\includeTikz{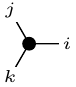}{}
\end{array}
&=\begin{cases}
1\qquad\text{if $i+j=k \mod{2}$}\\
0\qquad\text{otherwise}
\end{cases}.
\end{align}
Recall that our notation only indicates virtual indices, with a tensor at each vertex. The colors in the PEPS diagram of Eqn.~\ref{eqn:translationdefect} are included for comparison with the lattice diagram in Fig.~\ref{fig:tctranslationdefect} as described below. The presence of a nontrivial operation along the line where the MPOs were merged indicates the symmetry is not transversal.

The end points of this line correspond to lattice dislocations as shown in Fig.~\ref{fig:tctranslationdefect}a.  The state (\ref{eqn:translationdefect}) is the ground state of the toric code defined on this lattice. By performing unitary gates and adding/removing ancilla qubits along the defect line, the lattice geometry can be restored. We remark that this line has no effect on the anyons, and so corresponds to a topologically trivial symmetry.

In this section, we have identified the behavior of MPO domain walls corresponding to symmetry action on adjacent regions of the toric code. From this we observed that the physical APS action is not transversal (on-site), rather the merging of walls leads to a lattice dislocation.

\subsubsection{Symmetry action on excited states}

The domain wall constructed in Eqn.~\ref{eqn:TCWall} corresponds to the action of the symmetry on the vacuum. We want to understand how to transform other low energy states, namely those with anyons inserted. This will allow us to propagate the domain wall across the lattice. From a quantum computing perspective, this will allow us to understand the action of fault-tolerant logic gates on states with local errors. Since the APS action permutes the set of anyons, we expect the anyons in the interior of the domain to be transformed appropriately.

\begin{figure}
\centering
\includeTikz{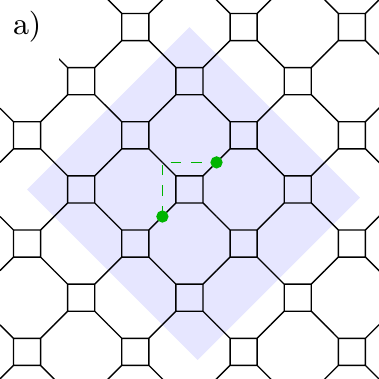}{}\hspace{5mm}
\includeTikz{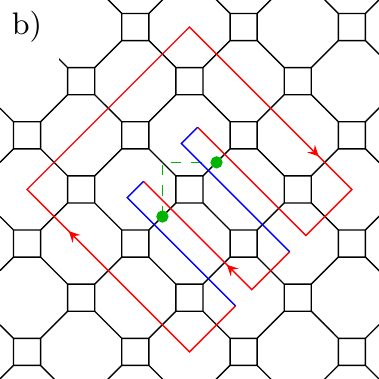}{}\\
\vspace{5mm}
\includeTikz{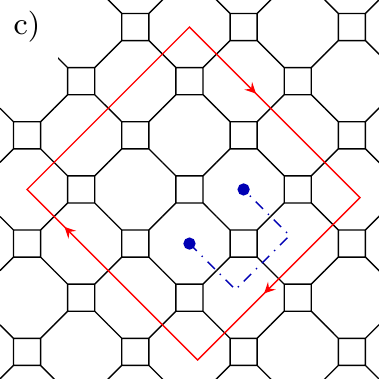}{}
\caption{To apply the APS to a region containing anyons, for example a pair of $e$ excitations (a), we need to deform the domain wall (using Eqn.~\ref{eqn:tcbubble}) so that the interior contains only vacuum (b). Once the domain wall MPO has been inserted, the anyon operators can be pushed through to the interior (c).}\label{fig:tcavoid}
\end{figure}

If there is an anyon within the region to which the APS is applied, as in Fig.~\ref{fig:tcavoid}a, there is an obstruction to placing the wall around the boundary. One can use the relation
\begin{align}
\begin{array}{c}
\includeTikz{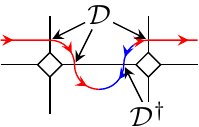}{}
\end{array}
&=
\begin{array}{c}
\includeTikz{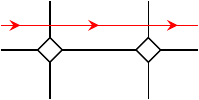}{}
\end{array}\label{eqn:tcbubble}
\end{align}
to deform the wall so as to avoid the excitation as shown in Fig.~\ref{fig:tcavoid}, leaving the interior in the vacuum. We remark that the meeting of red and blue lines corresponds to an identity tensor. The anyon string can now be commuted through the domain wall, leaving the transformed state in the interior.

\subsection{Symmetry defects and twists}

So far, we have discussed closed domain walls, corresponding to applying the APS to some region of the lattice. Given a symmetry, one can also consider inserting a defect line, which corresponds to allowing the domain wall to have open boundaries. Anyons crossing this line are transformed as usual, however we allow the line to terminate. We call this termination point a twist\cite{Bombin2010,Kitaev2012,Brown2013,Brown2016}, and it corresponds to a generalized topological charge\cite{Kitaev2012,Barkeshli2013,Barkeshli2014}.

The domain wall operator can be used to insert an APS defect into the PEPS
\begin{align}
\begin{array}{c}
\includeTikz{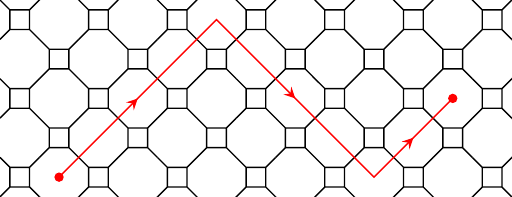}{}
\end{array}.
\end{align}

Since the MPO has a nontrivial bond dimension, merely inserting an open MPO into the PEPS would introduce additional physical (uncontracted) degrees of freedom. To retain the original spin lattice, appropriate boundary conditions for the MPO need to be chosen.
We can do this by considering the following process\cite{Bombin2010}. Create a pair of $e$ particles and move one of them around the end point. If we move it around once, the particle crosses the defect line once and becomes an $m$. Braiding it a second time recovers an $e$, which can be fused with the other particle
\begin{align}
\begin{array}{c}
\includeTikz{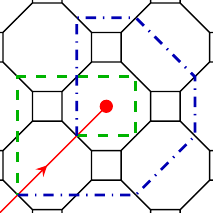}{}
\end{array}.\label{eqn:doublebraid}
\end{align}
Following \onlinecite{Bombin2010}, we define a generalized charge as a twist that is invariant under this process.

For the toric code, Eqn.~\ref{eqn:doublebraid}, reduces to
\begin{align}
\begin{array}{c}
\includeTikz{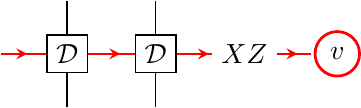}{}
\end{array}
&=\lambda
\begin{array}{c}
\includeTikz{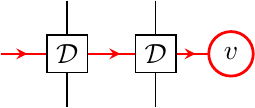}{}
\end{array}.
\end{align}
The ends should therefore be chosen to be eigenstates of $XZ$. We will refer to the twist with $v=|\pm i)$ as the end vector as $\sigma_{\pm}$ for consistency with \onlinecite{Bombin2010}. Choosing other end points corresponds to a superposition of $\sigma_+$ and $\sigma_-$.

We will now explore the topological properties of the twist defects resulting from this construction.

\subsubsection{Fusion}\label{S:fusion}

We can use the twist MPOs to compute the enriched fusion rules. Fusing a twist with an $m$ excitation
\begin{align}
\begin{array}{c}
\includeTikz{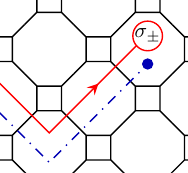}{}
\end{array}
&=
\begin{array}{c}
\includeTikz{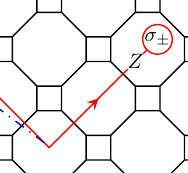}{}
\end{array}
=
\begin{array}{c}
\includeTikz{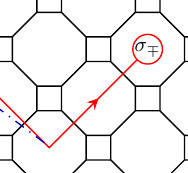}{}
\end{array},
\end{align}
changes the type of twist at the end point. The same is true when an $e$ is fused with a twist
\begin{align}
\begin{array}{c}
\includeTikz{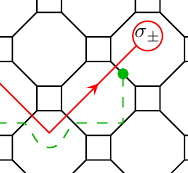}{}
\end{array}
&=
\begin{array}{c}
\includeTikz{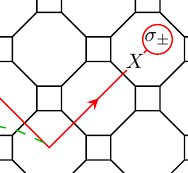}{}
\end{array}
=\pm i
\begin{array}{c}
\includeTikz{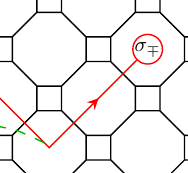}{}
\end{array}.
\end{align}

The situation is more complicated when two twists are fused. Following the discussion of domain wall mergers in Section~\ref{ss:growing}, we consider fusing a twist at the end of a $\D{}$ line with one that terminates a $\D{}^\dagger$ wall. If one instead attempted to fuse a pair of $\D{}$, there would be some topologically trivial transformation (corresponding to a lattice translation) which may conceal the nontrivial action. The result is
\begin{align}
\begin{array}{c}
\includeTikz{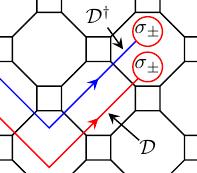}{}
\end{array}
&=
\begin{array}{c}
\includeTikz{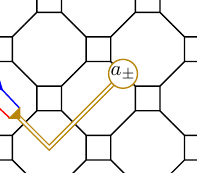}{}
\end{array},\\
\begin{array}{c}
\includeTikz{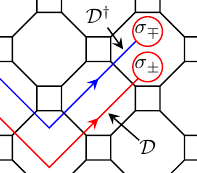}{}
\end{array}
&=
\begin{array}{c}
\includeTikz{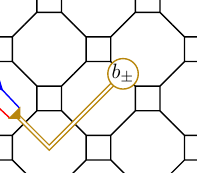}{}
\end{array},
\end{align}
where
\begin{align}
\begin{array}{c}
\includeTikz{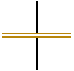}{}
\end{array}&=
\begin{pmatrix}
\one&0\\0&Z
\end{pmatrix}\\
\begin{array}{c}
\includeTikz{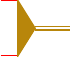}{}
\end{array}&=
\begin{pmatrix}
1&0\\0&1\\0&0\\0&0
\end{pmatrix}\\
\begin{array}{c}
\includeTikz{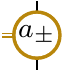}{}
\end{array}&=
\frac{1}{\sqrt{2}}\begin{pmatrix}
\one\\\pm i ZX
\end{pmatrix}\\
\begin{array}{c}
\includeTikz{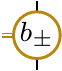}{}
\end{array}&=
\frac{1}{\sqrt{2}}\begin{pmatrix}
X\\\pm i Z
\end{pmatrix}.
\end{align}
The notation here identifies left (right) virtual indices to row (column) indices of the matrix and up (down) physical indices (black) correspond to operator indices of matrix entries.
Therefore, fusing $\sigma_\pm$ with $\sigma_\pm$ leaves a superposition of: a string of $\one$ terminated at the location of the tensor marked $a_\pm$ by a $\one$ (i.e. vacuum), and a string of $Z$ terminated by $ZX$ (i.e. an $m$ and $e$ at the same place, so therefore an $em$ particle). The fusion of $\sigma_\pm$ with $\sigma_\mp$ gives a superposition of: a string of $\one$ terminated by an $X$ (an $e$), and a string of $Z$ terminated by a $Z$ (an $m$).
The full set of fusion rules (neglecting phases) of the defects are therefore
\begin{align}
\sigma_\pm\times e&=\sigma_\mp&\sigma_\pm\times m&=\sigma_\mp\\
\sigma_\pm\times\sigma_\pm&=1+em&\sigma_\pm\times\sigma_\mp&=e+m.
\end{align}

These rules are consistent with the known rules for this enriched model\cite{Bombin2010}. This shows that the MPO construction gives an explicit realization of these defects.

\subsubsection{Hamiltonian terms for twists}\label{S:tcHams}
Using the domain wall/twist MPOs, one can construct Hamiltonians whose ground states correspond to the PEPS with twists inserted. One such set of Hamiltonian terms is
\begin{align}
S=\biggl\{
&\begin{array}{c}
\includeTikz{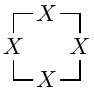}{}
\end{array},
\begin{array}{c}
\includeTikz{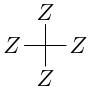}{}
\end{array},
\begin{array}{c}
\includeTikz{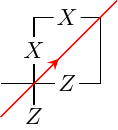}{}
\end{array},
\begin{array}{c}
\includeTikz{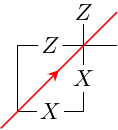}{}
\end{array},
\begin{array}{c}
\includeTikz{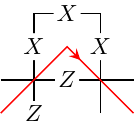}{}
\end{array},\nonumber\\
&
\begin{array}{c}
\includeTikz{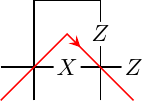}{}
\end{array},
\mp
\begin{array}{c}
\includeTikz{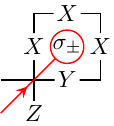}{}
\end{array},
\mp
\begin{array}{c}
\includeTikz{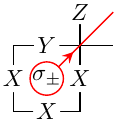}{}
\end{array}
\biggr\},\label{eqn:tctwistH}
\end{align}
where the final two terms correspond to the location of a twist of type $\sigma_\pm$.
By contracting these terms against the PEPS tensors, one can check that the PEPS is a $+1$ eigenstate of each. Since these terms are commuting stabilizers, the PEPS is the ground state of
$
H=-\sum_{h\in S} h.
$
We remark that this Hamiltonian was constructed without needing knowledge of the physical action of the APS on the underlying spin model, only the action on the anyon theory.
Since the ground states of the toric code with twist defects gives increased functionality for quantum computation compared with the bare model\cite{Brown2016}, it is important to know how to prepare such states. The parent Hamiltonian construction for PEPS\cite{Schuch2010a,Perez-Garcia2008} may provide a way to find such Hamiltonians away from fixed point models, where the physical APS action may be complicated.

\section{Color Code}\label{S:colorcode}

\begin{figure}
\centering
\includeTikz{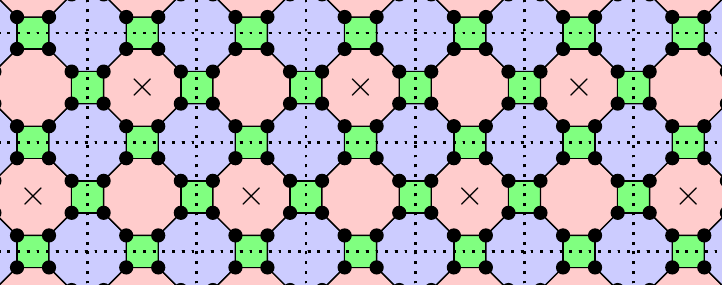}{}
\caption{The color code is defined on a three-colorable lattice. In this work, we consider the 4.8.8 lattice for simplicity. Qubits are located on the vertices. By applying a unitary transformation to each green square of qubits, the color code can be decoupled into a pair of toric codes (located on the dashed lattice) and a pair of local qubits. The PEPS tensors end up centered on the plaquettes indicated by `$\times$'.}\label{fig:488}
\end{figure}

As a second illustrative example, we now construct the domain wall MPOs for a model with a richer APS group than that of the toric code: the color code or $\ZZ_2\times\ZZ_2$ quantum double. This phase is particularly interesting from a quantum computing viewpoint, since the APS group is sufficiently rich that the full logical Clifford group can be implemented. This is the largest group of gates that can be fault-tolerantly implemented on a 2D qubit stabilizer or subsystem code\cite{Bravyi2013a,Pastawski2014}.


The color code is locally equivalent to two copies of the toric code\cite{Bombin2011,Kubica2015}. We will make extensive use of this equivalence.
The color code phase therefore has 16 abelian anyons, referred to as $e_T,\,e_B,\,m_T,\,m_B$ and their fusion products. The anyons labeled $T$ (top) and those labeled $B$ (bottom) independently generate two copies of the toric code particles.
The $\Tmat$ matrix of the theory is defined by $\Tmat_{e_im_j}=-\delta_{i,j}$, where $e_im_j:=e_i\times m_j$ and $\delta_{i,j}=1$ if $i=j$. The $\Smat$ matrix is defined by $\Smat_{e_i,m_j}=-\delta_{i,j}/4$. Unlike the toric code, the symmetry group of this theory is nonabelian, and has 72 elements. We will discuss the APS in detail in Section.~\ref{S:ccsymmetries}.

A stabilizer Hamiltonian that realizes the two dimensional color code can be defined on any three-colorable, three-valent lattice. In particular, we specify to the 4.8.8 lattice (\Fref{fig:488}) for simplicity. The Hamiltonian defining the code is
\begin{align}
H_{CC}&=-\sum_{p \in \mathrm{plaq.}}\left(\bar{X}_p+\bar{Z}_p\right),
\end{align}
where $\bar{P}_p$ indicates the tensor product of $P$ acting on all spins in plaquette $p$.

On the lattice, the equivalence to two copies of the toric code can be seen by acting with a local unitary circuit. This circuit acts independently on each green (square) plaquette. It is convenient to number the qubits
\begin{align}
\begin{array}{c}
\includeTikz{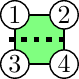}{}
\end{array}
&&
\begin{array}{c}
\includeTikz{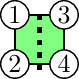}{}
\end{array},
\end{align}
where the dotted lines correspond to those in Fig.~\ref{fig:488}. The circuit then acts as
\begin{align}
\begin{array}{c}
\includeTikz{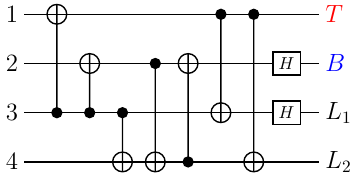}{}
\end{array},\label{eqn:tctocccircuit}
\end{align}
with the leftmost gate being applied first. In this transformed code, the `\textcolor{black}{$L$}' qubits are fully localized, with the ground state being $\ket{0}^{\otimes N}$, so can be discarded. On the remaining spins, the Hamiltonian corresponds to two independent toric codes, one defined on the `top' \textcolor{red}{$T$} qubits and the other on the `bottom' \textcolor{blue}{$B$} spins.
As mentioned above, the color code is interesting despite this equivalence as the full Clifford group can be implemented transversally. 

At all (green) plaquettes, the output qubits are labelled
\begin{align}
\begin{array}{c}
\includeTikz{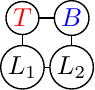}{}
\end{array}.
\end{align}

Following this transformation, the stabilizers of the color code become
\begin{align}
\bar{X}_R&\mapsto
\begin{array}{c}
\includeTikz{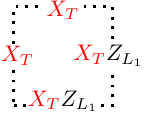}{}
\end{array}
&&\bar{Z}_R\mapsto
\begin{array}{c}
\includeTikz{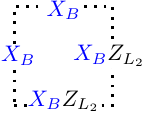}{}
\end{array}\\
\bar{X}_B&\mapsto
\begin{array}{c}
\includeTikz{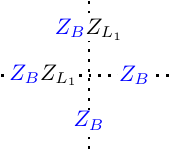}{}
\end{array}
&&\bar{Z}_B\mapsto
\begin{array}{c}
\includeTikz{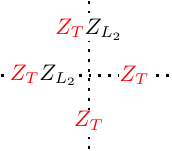}{}
\end{array}\\
\bar{X}_G&\mapsto\textcolor{black}{Z_{L_1}},\,
&&\bar{Z}_G\mapsto\textcolor{black}{Z_{L_2}},
\end{align}
where the new code is defined on the dotted lattice in Fig.~\ref{fig:488} with four spins per edge.
The PEPS tensor for this code is simply two copies of that in Eqn.~\ref{eqn:tcpepstensor}
\begin{align}
\begin{array}{c}
\includeTikz{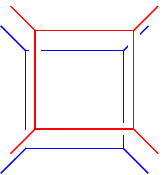}{}
\end{array}
:=
\begin{array}{c}
\includeTikz{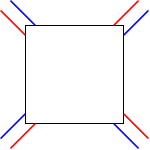}{}
\end{array}.
\end{align}
To construct the full PEPS tensor for the color code, the discarded \textcolor{black}{$L$} qubits need to be inserted and the (inverse of) the circuit Eqn.~\ref{eqn:tctocccircuit} applied.

\subsection{Excitations, anyon-permuting symmetries and domain wall operators}\label{S:ccsymmetries}

We now introduce a generating set of APS transformations for the color code anyons\cite{Yoshida2015a}, and then construct domain wall MPOs implementing the appropriate transformation. The color code supports four anyons and their fusion products. Generating anyons correspond to violating either a $Z$ or $X$ type plaquette of color red or blue. We will refer to these particles using these labels, for example the anyon corresponding to a violation of a red $Z$ type stabilizer will be labelled $r_z$, whilst $b_x$ will refer to a violation of a blue $X$ type plaquette. The green particles can be seen as a fusion of these, for example $g_z=r_z\times b_z$. We can use the circuit (\ref{eqn:tctocccircuit}) to find the toric code anyons corresponding to those in the color code
\begin{align}
r_x&\mapsto e_T&& b_z\mapsto m_T\\
r_z&\mapsto e_B&& b_x\mapsto m _B.
\end{align}

The permuting symmetries of these particles correspond to permutations of Pauli labels and exchanging Pauli labels and colors. Following the notation of \onlinecite{Yoshida2015a}, we label the symmetry elements as $W_i$. The APS group is generated by the 3-cycle $z\to x\to xz$
\begin{align}
&W_1:
\begin{tabular}{c c c p{5em} c c c}
\multicolumn{3}{c}{Color code}&&\multicolumn{3}{c}{Toric codes}\\
\hline
$r_z$&$\mapsto$&$r_x$&&$e_B$&$\mapsto$&$e_T$\\
$b_z$&$\mapsto$&$b_x$&&$m_T$&$\mapsto$&$m_B$\\
$r_x$&$\mapsto$&$r_xr_z$&&$e_T$&$\mapsto$&$e_Te_B$\\
$b_x$&$\mapsto$&$b_xb_z$&&$m_B$&$\mapsto$&$m_Tm_B$
\end{tabular},
\intertext{the 2-cycle $x\leftrightarrow z$}
&W_2:
\begin{tabular}{c c c p{6em} c c c}
$r_z$&$\mapsto$&$r_x$&&$e_B$&$\mapsto$&$e_T$\\
$b_z$&$\mapsto$&$b_x$&&$m_T$&$\mapsto$&$m_B$\\
$r_x$&$\mapsto$&$r_z$&&$e_T$&$\mapsto$&$e_B$\\
$b_x$&$\mapsto$&$b_z$&&$m_B$&$\mapsto$&$m_T$
\end{tabular},
\intertext{and the Pauli/color exchanging transformation $r\leftrightarrow x,\,b\leftrightarrow z$}
&W_5:
\begin{tabular}{c c c p{6em} c c c}
$r_z$&$\mapsto$&$b_x$&&$e_B$&$\mapsto$&$m_B$\\
$b_z$&$\mapsto$&$b_z$&&$m_T$&$\mapsto$&$m_T$\\
$r_x$&$\mapsto$&$r_x$&&$e_T$&$\mapsto$&$e_T$\\
$b_x$&$\mapsto$&$r_z$&&$m_B$&$\mapsto$&$e_B$
\end{tabular}.
\end{align}

We have already seen that the toric code supports an $e\leftrightarrow m$ APS. The color code therefore supports this symmetry too. Applying this duality to the bottom toric code corresponds to $W_5$. The transformation $W_2$ corresponds to swapping the two toric codes. The operator $W_1$ corresponds to a transversal controlled $X$ gate from the top toric code to the bottom, followed by a swap. It will be more convenient to work with the generators $\{\tilde{W}_1=W_2W_1,\,W_2,\,W_5\}$.

\subsubsection{Domain wall operators}

It is very straightforward to construct domain wall MPOs for the color code PEPS. The wall corresponding to $W_5$ is simply the one constructed in Eqn.~\ref{eqn:TCWall} on the second toric code, with identity on the first. The wall operator for $\tilde{W}_1$ is a controlled $X$ operator on each doubled bond, controlled on the top toric code. Finally, the wall for $W_2$ is a swap gate applied to each doubled bond.

\begin{align}
\tilde{W}_1&\to
\begin{array}{c}
\includeTikz{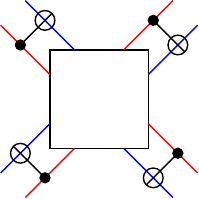}{}
\end{array}\label{eqn:w1}
,\\
W_2&\to
\begin{array}{c}
\includeTikz{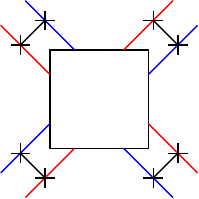}{}
\end{array}.
\end{align}

\subsection{Anyon-permuting symmetry twists for the color code}

The APS twists for $\tilde{W}_1$ and $W_2$ are straightforward to construct as the MPO already has trivial bond dimension and the twists correspond to definite generalized charges. Using the same arguments as in Sec.~\ref{S:tcHams}, one can check that the modified Hamiltonian terms for $\tilde{W}_1$ are
\begin{align}
\biggl\{&
\begin{array}{c}\scalebox{0.73}{
\includeTikz{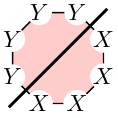}{}}
\end{array},
\begin{array}{c}\scalebox{0.73}{
\includeTikz{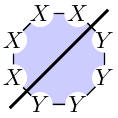}{}}
\end{array},
-\begin{array}{c}\scalebox{0.73}{
\includeTikz{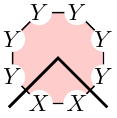}{}}
\end{array},
-\begin{array}{c}\scalebox{0.73}{
\includeTikz{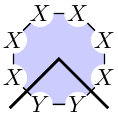}{}}
\end{array},
\begin{array}{c}\scalebox{0.73}{
\includeTikz{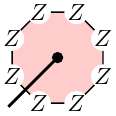}{}}
\end{array}
\biggr\},\label{eqn:w1twistH}
\end{align}
where the $X$ type term at the location of the twist is removed. For the twists associated with $W_2$, the modified terms are
\begin{align}
\biggl\{&
\begin{array}{c}\scalebox{0.73}{
\includeTikz{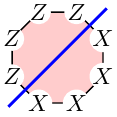}{}}
\end{array},
\begin{array}{c}\scalebox{0.73}{
\includeTikz{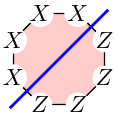}{}}
\end{array},
\begin{array}{c}\scalebox{0.73}{
\includeTikz{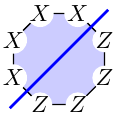}{}}
\end{array},
\begin{array}{c}\scalebox{0.73}{
\includeTikz{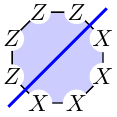}{}}
\end{array},
\begin{array}{c}\scalebox{0.73}{
\includeTikz{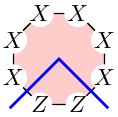}{}}
\end{array},
\nonumber\\
&
\begin{array}{c}\scalebox{0.73}{
\includeTikz{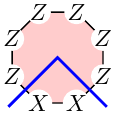}{}}
\end{array},
\begin{array}{c}\scalebox{0.73}{
\includeTikz{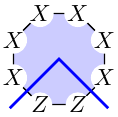}{}}
\end{array},
\begin{array}{c}\scalebox{0.73}{
\includeTikz{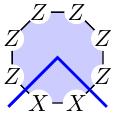}{}}
\end{array},
\begin{array}{c}\scalebox{0.73}{
\includeTikz{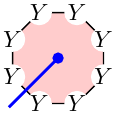}{}}
\end{array}
\biggr\},
\end{align}
where the $Y$ type operator at the twist location is the only term there.
For $W_5$, the modified terms are
\begin{align}
\biggl\{&
\begin{array}{c}\scalebox{0.73}{
\includeTikz{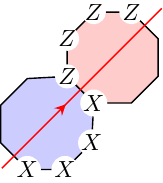}{}}
\end{array},
\begin{array}{c}\scalebox{0.73}{
\includeTikz{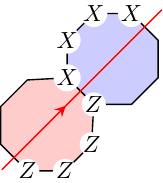}{}}
\end{array},
\begin{array}{c}\scalebox{0.73}{
\includeTikz{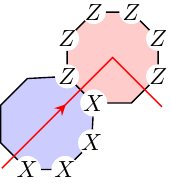}{}}
\end{array},
\begin{array}{c}\scalebox{0.73}{
\includeTikz{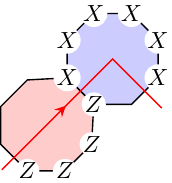}{}}
\end{array},
\nonumber\\
&
\pm
\begin{array}{c}\scalebox{0.73}{
\includeTikz{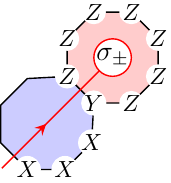}{}}
\end{array},
\pm
\begin{array}{c}\scalebox{0.73}{
\includeTikz{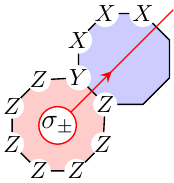}{}}
\end{array}\biggr\}.\label{eqn:cctwistH}
\end{align}

Although these Hamiltonians were not constructed using the usual parent Hamiltonian approach, we posit that such a construction could be used to find gapped Hamiltonian for more general APS twists. The Hamiltonian terms for these states at the level of the doubled toric code are provided in Appendix~\ref{S:twisthamiltonians}.

\subsection{Symmetry protected nature of the domain wall}\label{S:SPT}

It was noted in \onlinecite{Yoshida2015a} that the domain wall $\tilde{W}_1$ is associated with a one-dimensional Hamiltonian in a nontrivial $\ZZ_2\times\ZZ_2$ symmetry protected topological (SPT) phase. Given the MPOs we have constructed, we can explore this correspondence in the MPO framework. On the physical lattice, this required the definition of an `excitation basis'. Using this basis required care as a state corresponding to a single anyon is unphysical. In the framework of PEPS, this observation becomes straightforward.

Since the $\tilde{W}_1$ domain wall acts on the top toric code with only $\one$ and $Z$, it can only create $1$ and $m$ anyons. On the bottom toric code, only $1$ and $e$ particles can be created.

Looking at Eqn.~\ref{eqn:tcee}, we notice that the $e$ particle becomes `localized' on the PEPS. By this we mean that the occurrence of an $X$ operator on the virtual level signals the location of an excitation, with no string attached to it. When looking for the presence on an $m$ particle, it is not enough to look at a single bond since a $Z$ may be part of a string.
By modifying the PEPS, we can `localize' all particles to a single bond in this sense. The modified PEPS is
\begin{align}
\begin{array}{c}
\includeTikz{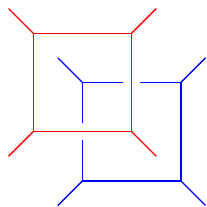}{}
\end{array}
:=
\begin{array}{c}
\includeTikz{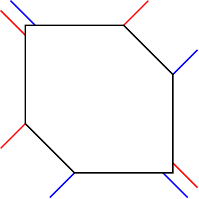}{}
\end{array},
\end{align}
where the top (red) PEPS tensor is that in Eqn.~\ref{eqn:tcpepstensor} with a Hadamard gate contracted onto each index, both virtual and physical. The bottom (blue) tensor is left unchanged. Red bonds of this new PEPS correspond to plaquettes in the top toric code. On this PEPS, $m_{T}$ excitations are created by inserting $X$ onto the appropriate red bond, whilst $m_{B}$ still corresponds to $X$ on the appropriate blue bond. There are no strings for either type of excitation, and in this way the existence of an $X$ directly corresponds to an excitation.

On this PEPS, we can write the $\tilde{W}_1$ wall as
\begin{align}
\begin{array}{c}
\includeTikz{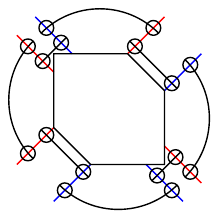}{}
\end{array},\label{eqn:sptcircuit}
\end{align}
where the virtual gate is a controlled phase gate in the $X$ basis. This circuit can be recognized as the one that creates the cluster state\cite{Briegel2000,Raussendorf2002} from the vacuum $\ket{0}^{\otimes N}$. The domain wall therefore has nontrivial SPT order with respect to a $\ZZ_2\times\ZZ_2$ symmetry. On this modified PEPS, the virtual symmetry identifying the topological phase is
\begin{align}
\begin{array}{c}
\includeTikz{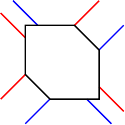}{}
\end{array}
&=
\begin{array}{c}
\includeTikz{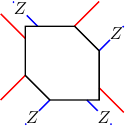}{}
\end{array}
=
\begin{array}{c}
\includeTikz{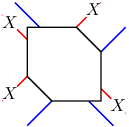}{}
\end{array},
\end{align}
which commutes with the circuit Eqn.~\ref{eqn:sptcircuit}. We therefore see that the SPT nature of this domain wall is protected by the virtual symmetry present throughout the phase, rather than being associated with any property of the stabilizer code.

A similar analysis could be performed for other walls, for example $W_2\tilde{W}_1 W_2$. In this case, we would also identify the wall as having an SPT property since the MPO is as in Eqn.~\ref{eqn:w1}, but with the control and target qubits exchanged. We believe that this framework of virtual MPOs provides a promising avenue to understanding the origin and nature of this SPT.

\section{Conclusions}\label{S:conclusions}

We have investigated the interplay between topological order and anyon-permuting symmetries in projected entangled pair states. By finding anyon-permuting domain walls, in the form of matrix product operators, we have realized the full APS in two models of interest.

Using these MPOs, we have shown how to introduce APS defect\add{s} (twists) into the PEPS by finding appropriate boundary conditions for the MPO. This allows the defect fusion rules to be obtained directly on the PEPS\add{, as discussed in Section~\ref{S:fusion}}. Further, Hamiltonians that realize the PEPS with twist insertions can easily be constructed.

The most obvious extension of this work is to the more general class of MPO-injective PEPS\cite{Buerschaper2014,Sahinoglu2014a,Bultinck2017}. This class realizes all known \add{non-chiral} topological orders, by using virtual MPO symmetries in place of the virtual group symmetry in Eqn.~\ref{eqn:ginj}. The examples we have discussed generalize straightforwardly to that framework, but a general condition for permuting anyons is unclear. In particular, we do not know how to formulate a local rule for finding an MPO tensor such that the full MPO performs the appropriate permutation of anyon sectors.

This work aimed to construct domain wall operators without otherwise altering the PEPS. In particular, we did not require that the APS acts in a transversal manner. Using a generalization of the string-net PEPS tensors\cite{Sahinoglu2014a,Bultinck2017}, one can construct states and MPO domain walls for symmetry enriched topological orders with an transversal symmetry action\cite{Williamson}. \onlinecite{Williamson} requires that the symmetry action can be made transversal. As a result of our work, we conjecture that the restriction to transversal APS is not required.

It would be interesting to deform the PEPS away from the fixed point and observe the effect on the domain wall operators. In particular, we expect the SPT discussed in Section~\ref{S:SPT} to be a property of the topological phase. It would be interesting to investigate the breakdown at the phase transition to a trivial phase.

One of the primary uses of PEPS is as a variational class for numerical optimization. The identification of domain wall MPOs in numerically obtained tensors would provide a way to identify the logical gates in models away from the fixed point.
\add{One would first need to obtain the MPOs that create anyons on the PEPS, possibly using a generalization of the algorithm introduced in \onlinecite{Bridgeman2016}. Given these anyon MPOs, one could attempt to find symmetry MPOs that permute them appropriately. We leave the design of such an algorithm to future work.}

The Hamiltonians in Eqns.~\ref{eqn:tctwistH}, \ref{eqn:w1twistH}-\ref{eqn:cctwistH} and Appendix~\ref{S:twisthamiltonians} did not make use of the parent Hamiltonian\cite{Schuch2010a,Perez-Garcia2008} construction. It may be possible to extend the proofs of gapped parents to the case of PEPS with MPO twist insertions, which would allow construction of gapped Hamiltonians for states with twist defects away from fixed point models.

In this paper, we have focused on the simplest topologically ordered models: the toric and color codes. The general approach is not limited to these models, and we see no obstruction to generalizing general quantum double models. In Appendix~\ref{S:quantumdoublemodels}, we construct domain walls for some of the symmetries of the cyclic quantum doubles. The extension to all abelian phases follows by stacking layers, in the same way that the color code is built from stacked toric codes. The extension to general groups is extremely interesting.

\acknowledgments
We acknowledge support from the Australian Research Council via the Centre of Excellence in Engineered Quantum Systems (EQuS), project number CE110001013, and
via project DP170103073. We thank Rafael Alexander, Daniel Barter, Benjamin Brown, Nick Bultinck, Christopher Chubb, Sam Elman, Steven Flammia, Robin Harper, Markus Kesselring, Sam Roberts, Thomas Smith and especially Dominic Williamson for useful discussions.

%

\onecolumngrid
\appendix

\clearpage
\section{Hamiltonians for twists}\label{S:twisthamiltonians}

In this appendix, we provide complete stabilizer Hamiltonians for the color code symmetry twisted states, both on the doubled toric code and color code level.

The symmetry twists for $\tilde{W}_1$ and $W_2$ are straightforward to construct as the MPO already has trivial bond dimension and the twists correspond to definite generalized charges. The toric code Hamiltonian for the $\tilde{W}_1$ twist is given by the stabilizers
\begin{align}
\biggl\{
&\begin{array}{c}\scalebox{0.82}{
\includeTikz{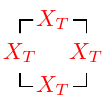}{}
}\end{array},
\begin{array}{c}\scalebox{0.82}{
\includeTikz{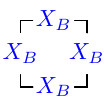}{}
}\end{array},
\begin{array}{c}\scalebox{0.82}{
\includeTikz{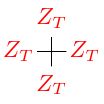}{}
}\end{array},
\begin{array}{c}\scalebox{0.82}{
\includeTikz{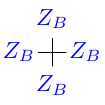}{}
}\end{array},
\begin{array}{c}\scalebox{0.82}{
\includeTikz{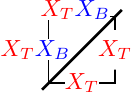}{}
}\end{array},
\begin{array}{c}\scalebox{0.82}{
\includeTikz{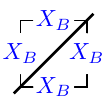}{}
}\end{array},
\begin{array}{c}\scalebox{0.82}{
\includeTikz{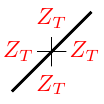}{}
}\end{array},
\begin{array}{c}\scalebox{0.82}{
\includeTikz{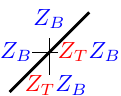}{}
}\end{array},
\begin{array}{c}\scalebox{0.82}{
\includeTikz{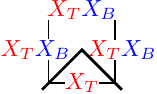}{}
}\end{array},
\begin{array}{c}\scalebox{0.82}{
\includeTikz{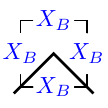}{}
}\end{array},
\begin{array}{c}\scalebox{0.82}{
\includeTikz{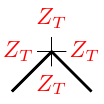}{}
}\end{array},
\nonumber\\
&
\begin{array}{c}\scalebox{0.82}{
\includeTikz{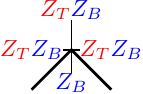}{}
}\end{array},
\begin{array}{c}\scalebox{0.82}{
\includeTikz{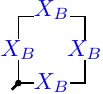}{}
}\end{array},
\begin{array}{c}\scalebox{0.82}{
\includeTikz{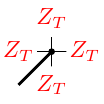}{}
}\end{array},
\begin{array}{c}\scalebox{0.82}{
\includeTikz{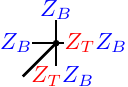}{}
}\end{array}
\biggr\},
\end{align}
and on the color code
\begin{align}
\biggl\{&
\begin{array}{c}\scalebox{0.82}{
\includeTikz{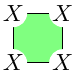}{}
}\end{array},
\begin{array}{c}\scalebox{0.82}{
\includeTikz{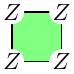}{}
}\end{array},
\begin{array}{c}\scalebox{0.82}{
\includeTikz{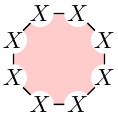}{}
}\end{array},
\begin{array}{c}\scalebox{0.82}{
\includeTikz{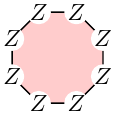}{}
}\end{array},
\begin{array}{c}\scalebox{0.82}{
\includeTikz{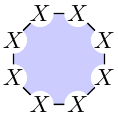}{}
}\end{array},
\begin{array}{c}\scalebox{0.82}{
\includeTikz{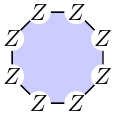}{}
}\end{array},
\begin{array}{c}\scalebox{0.82}{
\includeTikz{cctwistw1_red_B1}{}
}\end{array},
\begin{array}{c}\scalebox{0.82}{
\includeTikz{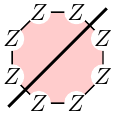}{}
}\end{array},
\begin{array}{c}\scalebox{0.82}{
\includeTikz{cctwistw1_blue_B1}{}
}\end{array},
\begin{array}{c}\scalebox{0.82}{
\includeTikz{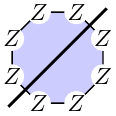}{}
}\end{array},
-\begin{array}{c}\scalebox{0.82}{
\includeTikz{cctwistw1_red_C1}{}
}\end{array},
\nonumber	\\
&
\begin{array}{c}\scalebox{0.82}{
\includeTikz{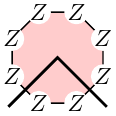}{}
}\end{array},
-\begin{array}{c}\scalebox{0.82}{
\includeTikz{cctwistw1_blue_C1}{}
}\end{array},
\begin{array}{c}\scalebox{0.82}{
\includeTikz{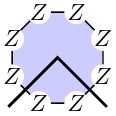}{}
}\end{array},
\begin{array}{c}\scalebox{0.82}{
\includeTikz{cctwistw1_red_D1}{}
}\end{array}
\biggr\}.
\end{align}

For $W_2$, the stabilizers are
\begin{align}
\biggl\{
&\begin{array}{c}\scalebox{0.82}{
\includeTikz{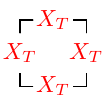}{}
}\end{array},
\begin{array}{c}\scalebox{0.82}{
\includeTikz{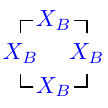}{}
}\end{array},
\begin{array}{c}\scalebox{0.82}{
\includeTikz{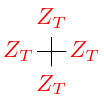}{}
}\end{array},
\begin{array}{c}\scalebox{0.82}{
\includeTikz{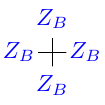}{}
}\end{array},
\begin{array}{c}\scalebox{0.82}{
\includeTikz{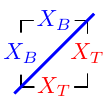}{}
}\end{array},
\begin{array}{c}\scalebox{0.82}{
\includeTikz{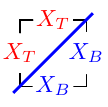}{}
}\end{array},
\begin{array}{c}\scalebox{0.82}{
\includeTikz{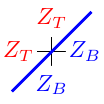}{}
}\end{array},
\begin{array}{c}\scalebox{0.82}{
\includeTikz{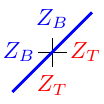}{}
}\end{array},
\begin{array}{c}\scalebox{0.82}{
\includeTikz{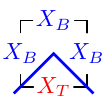}{}
}\end{array},
\begin{array}{c}\scalebox{0.82}{
\includeTikz{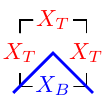}{}
}\end{array},
\begin{array}{c}\scalebox{0.82}{
\includeTikz{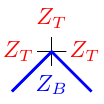}{}
}\end{array},
\nonumber\\
&
\begin{array}{c}\scalebox{0.82}{
\includeTikz{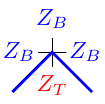}{}
}\end{array},
\begin{array}{c}\scalebox{0.82}{
\includeTikz{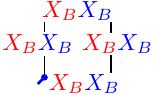}{}
}\end{array},
\begin{array}{c}\scalebox{0.82}{
\includeTikz{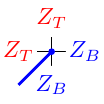}{}
}\end{array},
\begin{array}{c}\scalebox{0.82}{
\includeTikz{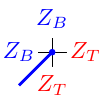}{}
}\end{array}
\biggr\}
\end{align}
on the toric code. On the color code, these become
\begin{align}
\biggl\{&
\begin{array}{c}\scalebox{0.82}{
\includeTikz{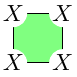}{}
}\end{array},
\begin{array}{c}\scalebox{0.82}{
\includeTikz{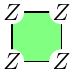}{}
}\end{array},
\begin{array}{c}\scalebox{0.82}{
\includeTikz{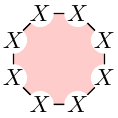}{}
}\end{array},
\begin{array}{c}\scalebox{0.82}{
\includeTikz{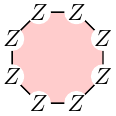}{}
}\end{array},
\begin{array}{c}\scalebox{0.82}{
\includeTikz{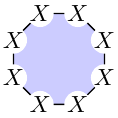}{}
}\end{array},
\begin{array}{c}\scalebox{0.82}{
\includeTikz{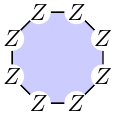}{}
}\end{array},
\begin{array}{c}\scalebox{0.82}{
\includeTikz{cctwistw2_red_B1}{}
}\end{array},
\begin{array}{c}\scalebox{0.82}{
\includeTikz{cctwistw2_red_B2}{}
}\end{array},
\begin{array}{c}\scalebox{0.82}{
\includeTikz{cctwistw2_blue_B1}{}
}\end{array},
\begin{array}{c}\scalebox{0.82}{
\includeTikz{cctwistw2_blue_B2}{}
}\end{array},
\begin{array}{c}\scalebox{0.82}{
\includeTikz{cctwistw2_red_C1}{}
}\end{array},
\nonumber\\
&
\begin{array}{c}\scalebox{0.82}{
\includeTikz{cctwistw2_red_C2}{}
}\end{array},
\begin{array}{c}\scalebox{0.82}{
\includeTikz{cctwistw2_blue_C1}{}
}\end{array},
\begin{array}{c}\scalebox{0.82}{
\includeTikz{cctwistw2_blue_C2}{}
}\end{array},
\begin{array}{c}\scalebox{0.82}{
\includeTikz{cctwistw2_red_D1}{}
}\end{array}
\biggr\}.
\end{align}

For $W_5$, the toric code terms are
\begin{align}
\biggl\{
&\begin{array}{c}\scalebox{0.82}{
\includeTikz{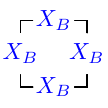}{}
}\end{array},
\begin{array}{c}\scalebox{0.82}{
\includeTikz{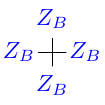}{}
}\end{array},
\begin{array}{c}\scalebox{0.82}{
\includeTikz{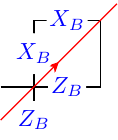}{}
}\end{array},
\begin{array}{c}\scalebox{0.82}{
\includeTikz{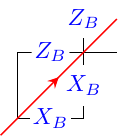}{}
}\end{array},
\begin{array}{c}\scalebox{0.82}{
\includeTikz{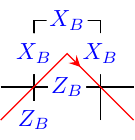}{}
}\end{array},
\begin{array}{c}\scalebox{0.82}{
\includeTikz{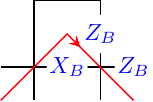}{}
}\end{array},
\mp
\begin{array}{c}\scalebox{0.82}{
\includeTikz{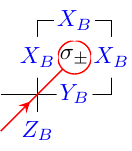}{}
}\end{array},
\mp
\begin{array}{c}\scalebox{0.82}{
\includeTikz{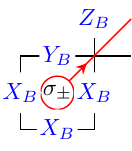}{}
}\end{array}
\biggr\},
\end{align}
along with all of the stabilizers of the top toric code, which remain unchanged. For the color code, we obtain
\begin{align}
\biggl\{&
\begin{array}{c}\scalebox{0.82}{
\includeTikz{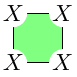}{}
}\end{array},
\begin{array}{c}\scalebox{0.82}{
\includeTikz{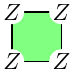}{}
}\end{array},
\begin{array}{c}\scalebox{0.82}{
\includeTikz{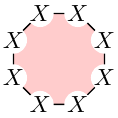}{}
}\end{array},
\begin{array}{c}\scalebox{0.82}{
\includeTikz{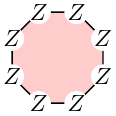}{}
}\end{array},
\begin{array}{c}\scalebox{0.82}{
\includeTikz{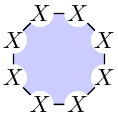}{}
}\end{array},
\begin{array}{c}\scalebox{0.82}{
\includeTikz{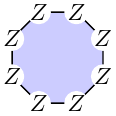}{}
}\end{array},
\begin{array}{c}\scalebox{0.82}{
\includeTikz{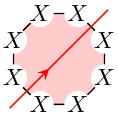}{}
}\end{array},
\begin{array}{c}\scalebox{0.82}{
\includeTikz{cctwistw5_red_B2}{}
}\end{array},
\begin{array}{c}\scalebox{0.82}{
\includeTikz{cctwistw5_blue_B1}{}
}\end{array},
\begin{array}{c}\scalebox{0.82}{
\includeTikz{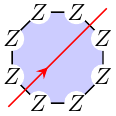}{}
}\end{array},
\begin{array}{c}\scalebox{0.82}{
\includeTikz{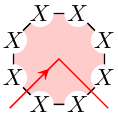}{}
}\end{array},
\nonumber\\
&
\begin{array}{c}\scalebox{0.82}{
\includeTikz{cctwistw5_red_C2}{}
}\end{array},
\begin{array}{c}\scalebox{0.82}{
\includeTikz{cctwistw5_blue_C1}{}
}\end{array},
\begin{array}{c}\scalebox{0.82}{
\includeTikz{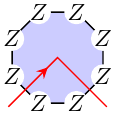}{}
}\end{array},
\begin{array}{c}\scalebox{0.82}{
\includeTikz{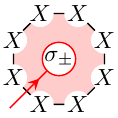}{}
}\end{array},
\pm
\begin{array}{c}\scalebox{0.82}{
\includeTikz{cctwistw5_red_D2}{}
}\end{array},
\begin{array}{c}\scalebox{0.82}{
\includeTikz{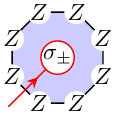}{}
}\end{array},
\begin{array}{c}\scalebox{0.82}{
\includeTikz{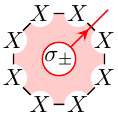}{}
}\end{array},
\pm
\begin{array}{c}\scalebox{0.82}{
\includeTikz{cctwistw5_blue_E1}{}
}\end{array},
\begin{array}{c}\scalebox{0.82}{
\includeTikz{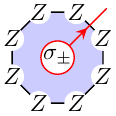}{}
}\end{array}\biggr\}.
\end{align}

\clearpage
\section{Abelian quantum double models}\label{S:quantumdoublemodels}

In this appendix, we generalize the domain walls to the case of the $\ZZ_N$ toric code\cite{Kitaev2003}. There are $N^2$ anyons in this theory which we label as $e^\alpha m^\beta$. The fusion rules for these particles are
\begin{align}
e^g m^\alpha\times e^h m^\beta&=e^{g+h}m^{\alpha+\beta},
\end{align}
where $+$ denotes addition mod $N$. The $\Tmat$ and $\Smat$ matrices are given by\cite{Coste2000,Coquereaux2012}
\begin{align}
\Tmat_{e^g m^\alpha}&=\chi^{\alpha}(g),\\
&=\omega^{\alpha g},\\
\Smat_{e^g m^\alpha,\,e^h m^\beta}&=\frac{1}{N^2}\left(\chi^\alpha(h)\chi^\beta(g)\right)^*,
\end{align}
where $\chi^i$ is the $i$th irreducible representation of $\ZZ_N$, $\omega=\exp(2\pi i/N)$, and $\cdot^*$ denotes complex conjugation.

The symmetry group of these particles is rather complicated\cite{Barkeshli2014,Lentner2017,Nikshych2013,Ostrik2003}, so we restrict to a $\ZZ_2\times\ZZ_2$ subgroup generated by
\begin{align}
\mathcal{D}:e^g m^\alpha &\mapsto e^\alpha m^g\\
\mathcal{C}:e^g m^\alpha &\mapsto e^{-g} m^{-\alpha},
\end{align}
which we will refer to as the duality and charge conjugation symmetry respectively. As we have discussed, this collapses to a $\ZZ_2$ when $N=2$ since $\mathcal{C}$ the particles are self inverse.

\subsection{Lattice Hamiltonian and PEPS}
The Hamiltonian for which Eqn.~\ref{eqn:tcpepstensor} is the ground state can be generalized to the $\ZZ_N$ topological order by replacing the two dimensional spins with $N$ dimensional ones.
Define the generalized Pauli operators so that
\begin{align}
Z\ket{j}&=\omega^j\ket{j}\\
X\ket{j}&=\ket{j-1},
\end{align}
where $j\in \ZZ_N$. The fixed point Hamiltonian we consider is
\begin{align}
H_{\ZZ_N}&=-\frac{1}{2}\sum \biggl(\begin{array}{c}
\includeTikz{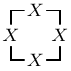}{}
\end{array}
+
\begin{array}{c}
\includeTikz{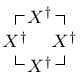}{}
\end{array}
\biggr)
-\frac{1}{2}\sum \biggl(\begin{array}{c}
\includeTikz{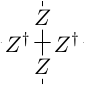}{}
\end{array}
+
\begin{array}{c}
\includeTikz{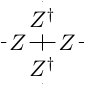}{}
\end{array}
\biggr)
.\label{eqn:znham}
\end{align}

Define the generalized Hadamard gate as
\begin{align}
H&=\frac{1}{\sqrt{N}}\sum_{j,k=0}^{N-1}\omega^{jk}\ketbra{j}{k},\label{eqn:znhad}
\end{align}
which acts on the Pauli operators as
\begin{align}
HXH^\dagger&=Z^\dagger\\
HZH^\dagger&=X.
\end{align}

One can readily verify that
\begin{align}
\begin{array}{c}
\includeTikz{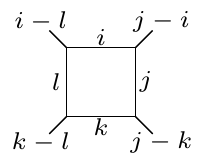}{}
\end{array}
=1,\label{eqn:znpepstensor}
\end{align}
defines a ground state for Eqn.~\ref{eqn:znham}.
This PEPS has a $\ZZ_N$ virtual symmetry generated by
\begin{align}
\begin{array}{c}
\includeTikz{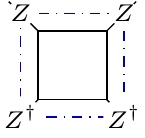}{}
\end{array}
&=
\begin{array}{c}
\includeTikz{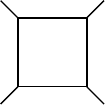}{}
\end{array}.\label{eqn:zntoposymmetry}
\end{align}

A state with an $m,\,m^{-1}$ pair is given by
\begin{align}
\begin{array}{c}
\includeTikz{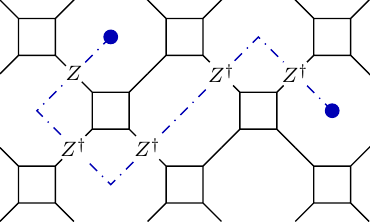}{}
\end{array},
\end{align}
where the path of the string is arbitrary since the virtual symmetry can be used to move it. A state with an $e,\,e^{-1}$ pair is created by
\begin{align}
\begin{array}{c}
\includeTikz{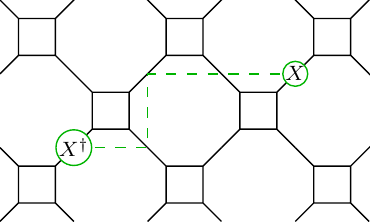}{}
\end{array}.\label{eqn:znee}
\end{align}

\subsection{Domain wall operators}

We begin with the MPO for the $\mathcal{C}$ symmetry, which implements charge conjugation on both $e$ and $m$ anyons. On the level of the virtual strings, this can be implemented using
\begin{align}
\begin{array}{c}
\includeTikz{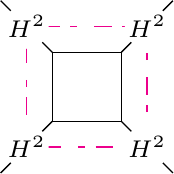}{}
\end{array},
\end{align}
where
\begin{align}
H^2&=\sum_{j=0}^{N-1}\ketbra{j}{-j}.
\end{align}
is the square of the $\ZZ_N$ Hadamard operator (Eqn.~\ref{eqn:znhad}).

The domain wall constructed for the $\ZZ_2$ case in Eqn.~\ref{eqn:TCWall} can be generalized to the $\ZZ_N$ case. Since the symmetry generator is no longer self-inverse, we need two MPO tensors
\begin{align}
\begin{array}{c}
\includeTikz{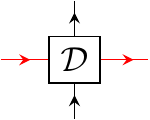}{}
\end{array}
&=\sqrt{N}\times
\begin{array}{c}
\includeTikz{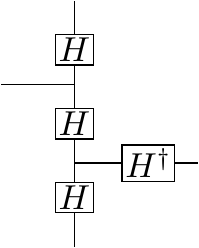}{}
\end{array}\label{eqn:znWalla}\\
\begin{array}{c}
\includeTikz{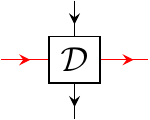}{}
\end{array}
&=\sqrt{N}\times
\begin{array}{c}
\includeTikz{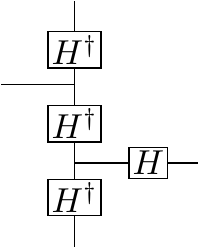}{}
\end{array},\label{eqn:znWallb}
\end{align}
where the PEPS tensors are dressed with arrows
\begin{align}
\begin{array}{c}
\includeTikz{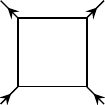}{}
\end{array},
\end{align}
indicating which bonds the symmetry in Eqn.~\ref{eqn:zntoposymmetry} acts as $Z$ (outgoing) and which it acts as $Z^\dagger$ (ingoing).

Definite symmetry twists can be found by eigenstates of the double braiding process depicted in Eqn.~\ref{eqn:doublebraid}, which results in eigenvectors of $Z^\dagger X$ being used to close the MPOs. There are $N$ distinct $\mathcal{D}$ twists, differing by the absorption of an $e$ or $m$ particle: $\sigma_{j}\times e=\sigma_{j+1}$, $\sigma_{j}\times m=\sigma_{j+1}$.

\subsection{Prime dimension codes}

For the special case of $\ZZ_p$ toric codes, with $p$ prime, we can characterize the full symmetry group. Consider the transformations
\begin{align}
\mathcal{D}:e^g m^\alpha &\mapsto e^\alpha m^g\\
\mathcal{Q}_n:e^g m^\alpha &\mapsto e^{n g} m^{\alpha i_n},
\end{align}
where $i_n$ is the \emph{modular inverse} of $n$ so $i_n\cdot n=1\mod p$. These transformations define a group $\G=\ZZ_p^\times\rtimes \ZZ_2\cong \Dih{p-1}$ of order $2(p-1)$, where $\ZZ_p^\times$ is the multiplicative group of integers modulo $p$, $\mathcal{D}\mathcal{Q}_n\mathcal{D}=\mathcal{Q}_{i_n}$ and $\Dih{n}$ is the dihedral group of order 2n.

One can check that the $\Smat$ and $\Tmat$ are preserved by this group. The symmetry $\mathcal{C}$ described above corresponds to $\mathcal{Q}_{p-1}$.

The domain wall MPO for the transformation $Q_n$ is described by the matrix
\begin{align}
\mathcal{Q}_n:=\sum_{j=0}^{p-1} \ketbra{n j}{j},
\end{align}
where $nj$ is taken modulo $p$.

\end{document}